\documentclass[aps,showpacs]{revtex4}
\usepackage{graphicx}
\usepackage{bm}
\begin{document}
\title{Density-matrix formalism with three-body ground-state correlations}
\author{Mitsuru Tohyama}
\affiliation{Kyorin University School of Medicine, Mitaka, Tokyo
  181-8611, Japan}
\author{Peter Schuck} 
\affiliation{Institut de Physique Nucl$\acute{\rm e}$aire, 
IN2P3-CNRS, Universit$\acute{\rm e}$ Paris-Sud, F-91406 Orsay Cedex, France} 
\date{\today}
\begin{abstract}
A density-matrix formalism which includes the effects of three-body ground-
state correlations 
is applied to the standard Lipkin model. The reason to consider the 
complicated three-body correlations is that the truncation scheme of reduced 
density matrices up to the two-body level does not give satisfactory results
to the standard Lipkin model.
It is shown that inclusion of the three-body correlations drastically 
improves the properties of the ground states and excited states. 
It is pointed out that lack of mean-field effects in the standard Lipkin model enhances 
the relative importance of the three-body ground-state correlations. 
Formal aspects of the density-matrix formalism such as a relation to the
variational principle and the stability condition of the ground state are also
discussed. It is pointed out that the three-body ground-state correlations are
necessary to satisfy the stability condition. 
\end{abstract}
\pacs{21.60.Jz}
\maketitle
\section{Introduction}
Mean-field theories such as the Hartree-Fock theory (HF), the Haree-Fock 
Bogoliubov theory, 
the random-phase approximation (RPA), and the quasi-particle RPA have
extensively been used to study the ground states and collective excitations 
of atomic nuclei \cite{RS}.
For more realistic theoretical treatment of nuclei such as inclusion of the ground-state correlations other
than pairing correlations and the damping effects of collective excitations, however,
we must go beyond the mean-field theories.
The time-dependent density-matrix theory (TDDM) \cite{WC,GT,Peter}
which has been formulated by truncating the chain of the equations of motion
for reduced density matrices up to the two-body level is one of such 
extended mean-field theories.
It has been pointed out that a stationary solution of the TDDM equations 
gives a correlated ground state
and that the small amplitude limit of TDDM based on the correlated ground state 
corresponds to an extended RPA (ERPA) including two-body transition 
amplitudes \cite{TG89}.  
To test the reliability of the density-matrix formalism, 
we have applied it to solvable models \cite{Lip,Yang,Taka03}
and found that the obtained results improve on deficiencies of the mean-field 
theories \cite{Takahara,mt07}.
In the case of the standard Lipkin model \cite{Lip} where the interaction 
term contains only two particle - two hole excitations, however,
we found that the ground states in TDDM become slightly overbound as compared 
with the exact solutions.
That the approximate ground states have lower energy than the exact ones 
contradicts the variational
principle \cite{RS} for the total wavefunction and should possibly be avoided, 
although TDDM is not based on a Raleigh-Ritz variational principle.
The aim of this paper is to clarify the origin of such an unsatisfactory 
feature of TDDM in the standard
Lipkin model. We will show that inclusion of three-body ground-state 
correlations, though it is complicated, drastically improves
the results for the standard Lipkin model.
The paper is organized as follows: The formulation of TDDM  
with the three-body ground-state correlations \cite{ts07,ts08} and ERPA built
on the TDDM ground state are given in sect.2. 
Some formal aspects of TDDM and ERPA, a relation of the TDDM equations to the 
variational principle, 
the stability condition of the ground-state, and  
a comparison of ERPA with the self-consistent RPA (SCRPA) \cite{Rowe,Dukelsky},
which have not been pointed out in our earlier publications,
are also discussed in sect. 2.
The results for the standard
Lipkin model are presented in sect.3. Section 4 is devoted to the summary.
\section{Formulation \label{sect.2}}
Although the numerical calculations are performed for the Lipkin model, the formulation  
is presented using 
the following general hamiltonian
\begin{eqnarray}
\hat{H}=\sum_{\lambda\lambda'} \langle\lambda|t|\lambda'\rangle a^{\dag}_\lambda a_{\lambda'}
+\frac{1}{2}\sum_{\lambda_1\lambda_2\lambda_1'\lambda_2'} \langle\lambda_1\lambda_2|v|\lambda_1'\lambda_2'\rangle a^{\dag}_{\lambda_1}a^{\dag}_{\lambda_2}
a_{\lambda_2'}a_{\lambda_1'},
\end{eqnarray}
where $t$ is the kinetic energy operator, $v$ is a two-body interaction
and $a^\dag_\lambda(a_\lambda)$ the creation (annihilation) operator 
of a nucleon in a single-particle state $\lambda$.
\subsection{TDDM}
In TDDM the ground state $|0\rangle$ is defined by the occupation matrix $n_{\alpha\alpha'}$, the two-body correlation matrix
$C_{\alpha\beta\alpha'\beta'}$ and the three-body correlation matrix
$C_{\alpha\beta\gamma\alpha'\beta'\gamma'}$
given by
\begin{eqnarray}
n_{\alpha\alpha'}&=&\langle0|a^{\dag}_{\alpha'}a_{\alpha}|0\rangle, \\
C_{\alpha\beta\alpha'\beta'}&=&\langle0|a^{\dag}_{\alpha'}a^{\dag}_{\beta'}a_{\beta}a_{\alpha}|0\rangle
-{\cal A}(n_{\alpha\alpha'}n_{\beta\beta'}), 
\label{C2}\\
C_{\alpha\beta\gamma\alpha'\beta'\gamma'}&=&
\langle0|a^{\dag}_{\alpha'}a^{\dag}_{\beta'}a^{\dag}_{\gamma'}a_{\gamma}a_{\beta}a_{\alpha}|0\rangle
-{\cal A}(n_{\alpha\alpha'}n_{\beta\beta'}n_{\gamma\gamma'}+{\cal S}(n_{\alpha\alpha'}C_{\beta\gamma\beta'\gamma'})),
\label{C3}
\end{eqnarray}
where ${\cal A}$ and ${\cal S}$ mean that the products in the parentheses
are properly antisymmetrized and symmetrized under 
the exchange of single-particle indices \cite{WC}.
Equations of motion for $n_{\alpha\alpha'}$, 
$C_{\alpha\beta\alpha'\beta'}$ and 
$C_{\alpha\beta\gamma\alpha'\beta'\gamma'}$ can be obtained by truncating a coupled 
chain of equations
of motion for reduced density matrices (the so-called Bogoliubov-Born-Green-
Kirkwood-Yvon hierarchy) 
up to the three-body level \cite{WC}
and can be written as
\begin{eqnarray}
i\hbar\frac{d}{dt}n_{\alpha\alpha'}&=&\langle0|[a^{\dag}_{\alpha'}a_{\alpha},\hat{H}]|0\rangle=F_1(\alpha\alpha'), 
\label{td1}
\\
i\hbar\frac{d}{dt}C_{\alpha\beta\alpha'\beta'}
&=&:\langle0|[a^{\dag}_{\alpha'}a^{\dag}_{\beta'}a_{\beta}a_{\alpha},\hat{H}]|0\rangle:=F_2(\alpha\beta\alpha'\beta'), 
\label{td2}
\\
i\hbar\frac{d}{dt}C_{\alpha\beta\gamma\alpha'\beta'\gamma'}&=&
:\langle0|[a^{\dag}_{\alpha'}a^{\dag}_{\beta'}a^{\dag}_{\gamma'}a_{\gamma}a_{\beta}a_{\alpha},\hat{H}]|0\rangle:
=F_3(\alpha\beta\gamma\alpha'\beta'\gamma'),
\label{td3}
\end{eqnarray}
where $:~:$ means that the time derivatives of the second terms on the right-hand sides of eqs. (\ref{C2}) and (\ref{C3}) are
subtracted.
Since a four-body correlation matrix is neglected, the expectation values of four-body operators in eq.(\ref{td3}) are approximated by the products of 
$n_{\alpha\alpha'}$, 
$C_{\alpha\beta\alpha'\beta'}$ and 
$C_{\alpha\beta\gamma\alpha'\beta'\gamma'}$.
The expressions 
for $F_1$, $F_2$ and $F_3$ are
given in the Appendix, where the single-particle states which satisfy the 
HF-like mean-field equation
\begin{eqnarray}
h(\rho)\phi_\alpha(1)=\epsilon_\alpha\phi_\alpha(1)
\label{hfeq}
\end{eqnarray}
are used.
Here, $\rho$ is the one-body density matrix given by 
$\rho(1,1')=\sum_{\alpha\alpha'}n_{\alpha\alpha'}\phi_\alpha(1)\phi^*_{\alpha'}(1')$ and numbers 
indicate spatial, spin and isospin coordinates.

To obtain the ground state implies that all quantities $n_{\alpha\alpha'}$, $C_{\alpha\beta\alpha'\beta'}$, 
$C_{\alpha\beta\gamma\alpha'\beta'\gamma'}$
and $\phi_\alpha$ are determined under the stationary conditions 
\begin{eqnarray}
i\hbar\frac{d}{dt}n_{\alpha\alpha'}&=&F_1(\alpha\alpha')=0, 
\label{gs1}
\\
i\hbar\frac{d}{dt}C_{\alpha\beta\alpha'\beta'}&=&F_2(\alpha\beta\alpha'\beta')=0, 
\label{gs2}
\\
i\hbar\frac{d}{dt}C_{\alpha\beta\gamma\alpha'\beta'\gamma'}&=&F_3(\alpha\beta\gamma\alpha'\beta'\gamma')=0.
\label{gs3}
\end{eqnarray}
and eq. (\ref{hfeq}).
This task can be achieved  
using the gradient method \cite{TTS}. The functional derivatives of eqs. (\ref{gs1})-(\ref{gs3}) are written as
\begin{eqnarray}
\left(
\begin{array}{ccc}
\frac{\delta F_1}{\delta n} &\frac{\delta F_1}{\delta C_2} & 0 \\
\frac{\delta F_2}{\delta n} & \frac{\delta F_2}{\delta C_2} & \frac{\delta F_2}{\delta C_3} \\
\frac{\delta F_3}{\delta n} & \frac{\delta F_3}{\delta C_2} & \frac{\delta F_3}{\delta C_3}
\end{array}
\right)
\left(
\begin{array}{c}
\Delta n\\
\Delta C_2\\
\Delta C_3
\end{array}
\right)
=
\left(
\begin{array}{c}
\Delta F_{1}\\
\Delta F_{2}\\
\Delta F_{3}
\end{array}
\right)=-
\left(
\begin{array}{c}
 F_{1}\\
 F_{2}\\
 F_{3}
\end{array}
\right).
\label{gradient0}
\end{eqnarray}
The inversion of the above matrix gives the following equation to be used in the gradient method
\begin{eqnarray}
\left(
\begin{array}{c}
n(N+1)\\
C_2(N+1)\\
C_3(N+1)
\end{array}
\right)
=
\left(
\begin{array}{c}
n(N)\\
C_2(N)\\
C_3(N)
\end{array}
\right)
-\alpha\left(
\begin{array}{ccc}
a &c & 0 \\
b& d& e \\
f & g & h
\end{array}
\right)^{-1}
\left(
\begin{array}{c}
F_{1}(N)\\
F_{2}(N)\\
F_{3}(N)
\end{array}
\right)
\label{gradient},
\end{eqnarray}
where $n(N)$, $C_2(N)$ and $C_3(N)$ imply
$n_{\alpha\alpha'}$, $C_{\alpha\beta\alpha'\beta'}$ and 
$C_{\alpha\beta\gamma\alpha'\beta'\gamma'}$ at the $N$th iteration step, respectively, and
$a=\delta F_1/ \delta n$, $b=\delta F_2/ \delta n$, $c=\delta F_1/ \delta C_2$,
$d=\delta F_2/ \delta C_2$, $e=\delta F_2/ \delta C_3$,
$f=\delta F_3/ \delta n$,
$g=\delta F_3/ \delta C_2$, and $h=\delta F_3/ \delta C_3$.
The expressions for these matrices are given in ref. \cite{ts08}.
The matrices $a$, $b$, $d$, $f$, $g$, and $h$ depend on the iteration step $N$ through $n(N)$, $C_2(N)$ and $C_3(N)$.
To solve eq. (\ref{gradient}), we start from a simple ground state such as the 
HF ground state where $n_{\alpha\alpha'}$, 
$C_{\alpha\beta\alpha'\beta'}$ and $C_{\alpha\beta\gamma\alpha'\beta'\gamma'}$ can be easily evaluated and iterate eq. (\ref{gradient})
until convergence is achieved.  A small parameter $\alpha$ is introduced to regulate the convergence process. 
When the mean-field potential is present, eq. (\ref{gradient}) couples to eq. (\ref{hfeq}) 
through $n_{\alpha\alpha'}$.
As was discussed in ref. \cite{ts08},
the matrix consisting of the functional derivatives of 
$F_1, ~F_2$ and $F_3$ 
on the right-hand side of eq. (\ref{gradient}) can be derived as the small amplitude limit of
TDDM and has a close relation with the hamiltonian matrix of 
ERPA given in the next subsection \cite{ts07,TS}. 
This indicates that the ground state is not independent of the excited states in ERPA. Let us discuss this point 
in some more detail using the eigenstates of the following matrix equation,
which corresponds to the small amplitude limit of eqs. (\ref{td1})-(\ref{td3}) \cite{ts08},
\begin{eqnarray}
\left(
\begin{array}{ccc}
a &c & 0 \\
b& d& e \\
f & g & h
\end{array}
\right)
\left(
\begin{array}{c}
x_\mu\\
y_\mu\\
z_\mu
\end{array}
\right)=\Omega_\mu\left(
\begin{array}{c}
x_\mu\\
y_\mu\\
z_\mu
\end{array}
\right).
\label{stddm}
\end{eqnarray}
Since the closure relation is given as \cite{TS},
\begin{eqnarray}
\sum_\mu\left(
\begin{array}{c}
x_\mu \\
y_\mu\\
z_\mu
\end{array}
\right)\left(\tilde{x}_\mu^*~\tilde{y}_\mu^*~\tilde{z}_\mu^*\right)=I,
\end{eqnarray}
where $(\tilde{x}_\mu^*~\tilde{y}_\mu^*~\tilde{z}_\mu^*)$ is the left-hand eigenvector of eq. (\ref{stddm})
and $I$ is unit matrix, eq. (\ref{gradient}) is written as
\begin{eqnarray}
\left(
\begin{array}{c}
n(N+1)\\
C_2(N+1)\\
C_3(N+1)
\end{array}
\right)
=
\left(
\begin{array}{c}
n(N)\\
C_2(N)\\
C_3(N)
\end{array}
\right)
-\alpha\sum_{\Omega_\mu\neq 0}\frac{1}{\Omega_\mu}\left(
\begin{array}{c}
x_\mu \\
y_\mu\\
z_\mu
\end{array}
\right)\left(\tilde{x}_\mu^*~\tilde{y}_\mu^*~\tilde{z}_\mu^*\right)
\left(
\begin{array}{c}
F_{1}(N)\\
F_{2}(N)\\
F_{3}(N)
\end{array}
\right).
\label{gradient1}
\end{eqnarray}
In eq. (\ref{gradient1}) only the eigenstates with $\Omega_\mu\neq 0$ can contribute as is understood by multiplying 
eq. (\ref{gradient0}) with $\left(\tilde{x}_\mu^*~\tilde{y}_\mu^*~\tilde{z}_\mu^*\right)$. 
The above equation indicates that the occupation matrix and the correlation matrices are constructed from the eigenvectors of 
eq. (\ref{stddm}) at each iteration step which have the same quantum numbers as the ground state: In the standard Lipkin model these
states are two-phonon states.

For the numerical calculations shown below we use not eq. (\ref{gradient1}) but eq. (\ref{gradient})
because a considerable dimension size of the three-body part of eq. (\ref{stddm})
makes it impracticable to adopt eq. (\ref{gradient1}) which
requires the eigenvalues and eigenvectors at each iteration step. In the matrix inversion of eq. (\ref{gradient}), 
however, mixing of unphysical states with 
$\Omega_\mu\approx0$ is unavoidable. This problem can be removed by slightly shifting the unperturbed energies in $a$, $d$, and $h$
by $\Delta \epsilon$ so that the inverse can be taken. 
In the case of the Lipkin model the typical value of $\Delta \epsilon$ used is $10^{-5}\epsilon$ where $\epsilon$ is
the level spacing of the Lipkin model hamiltonian. The obtained results are independent of $\Delta \epsilon$ because the product
\begin{eqnarray}
\left(\tilde{x}_\mu^*~\tilde{y}_\mu^*~\tilde{z}_\mu^*\right)
\left(
\begin{array}{c}
F_{1}(N)\\
F_{2}(N)\\
F_{3}(N)
\end{array}
\right)
\end{eqnarray} 
vanishes for the states with $\Omega_\mu=0$.
Since the inversion of the matrix on the right-hand side of eq. (\ref{gradient}) is
time consuming, the gradient method is supplemented with a time-dependent approach as will be explained below.
We make a comparison with a simplified version of TDDM where the three-body correlation matrix $C_{\alpha\beta\gamma\alpha'\beta'\gamma'}$
is neglected. We refer this TDDM as to TDDM1. 

\subsection{Extended RPA}
In this section we present our extended version of RPA (ERPA) \cite{ts08}. We also discuss its close connection to TDDM. ERPA is formulated for
the following excitation operator consisting of one-body and two-body operators
\begin{eqnarray}
\hat{Q}^{\dag}_{\mu}=\sum_{\lambda\lambda'}{x^\mu_{\lambda\lambda'}:a^{\dag}_\lambda a_{\lambda'}:}
+\sum_{\lambda_1\lambda_2\lambda_1'\lambda_2'}{X^\mu_{\lambda_1\lambda_2\lambda_1'\lambda_2'}:a^{\dag}_{\lambda_1}a^{\dag}_{\lambda_2}a_{\lambda_2'}a_{\lambda_1'}:},
\label{exoper}
\end{eqnarray}
where :~: implies that uncorrelated parts consisting of lower-level operators are to be subtracted; for example, 
\begin{eqnarray}
:a^{\dag}_{\alpha'}a_{\alpha}:&=&a^{\dag}_{\alpha'}a_{\alpha}-n_{\alpha\alpha'},
\\:a^{\dag}_{\alpha'}a^{\dag}_{\beta'}a_{\beta}a_{\alpha}:&=&a^{\dag}_{\alpha'}a^{\dag}_{\beta'}a_{\beta}a_{\alpha}
-{\cal AS}(n_{\alpha\alpha'}:a^{\dag}_{\beta'}a_{\beta}:)
\nonumber \\
&-&[{\cal A}(n_{\alpha\alpha'}n_{\beta\beta'})
+C_{\alpha\beta\alpha'\beta'}].
\end{eqnarray} 
It is assumed that the operator $\hat{Q}^{\dag}_\mu$ satisfies
\begin{eqnarray}
\hat{Q}^\dag_\mu|\Psi_0\rangle=|\Psi_\mu\rangle,\\
\hat{Q}_\mu|\Psi_0\rangle=0.
\label{scrpa} 
\end{eqnarray}
Here, $|\Psi_0\rangle$ is the ground state in ERPA and $|\Psi_\mu\rangle$ is an 
excited state. 
The ERPA equations are obtained from the equations-of-motion method
\cite{Rowe} 
\begin{eqnarray}
\langle\Psi_0|[[:a^{\dag}_{\alpha'}a_{\alpha}:,\hat{H}],\hat{Q}^{\dag}_{\mu}]|\Psi_0\rangle 
&=&\omega_\mu\langle\Psi_0|[:a^{\dag}_{\alpha'}a_{\alpha}:,\hat{Q}^{\dag}_{\mu}]|\Psi_0\rangle, 
\label{erpa1}\\
\langle\Psi_0|[[:a^{\dag}_{\alpha'}a^{\dag}_{\beta'}a_{\beta}a_{\alpha}:,\hat{H}],\hat{Q}^{\dag}_{\mu}]|\Psi_0\rangle 
&=&\omega_\mu\langle\Psi_0|[:a^{\dag}_{\alpha'}a^{\dag}_{\beta'}a_{\beta}a_{\alpha}:,\hat{Q}^{\dag}_{\mu}]|\Psi_0\rangle,
\label{erpa2}
\end{eqnarray}
where $\omega_\mu$ is the excitation energy of $|\Psi_\mu\rangle$.
In the evaluation of the matrix elements we approximate
the ERPA ground state $|\Psi_0\rangle$ by $|0\rangle$ which satisfies eqs. (\ref{hfeq})-(\ref{gs3}). 
The ERPA equations can then be written in matrix form
\begin{eqnarray}
\left(
\begin{array}{cc}
A & C \\
B & D 
\end{array}
\right)\left(
\begin{array}{c}
x^\mu \\
X^\mu
\end{array}
\right)=\omega_\mu\left(
\begin{array}{cc}
S_1 & T_1 \\
T_2 & S_2
\end{array}
\right)\left(
\begin{array}{c}
x^\mu \\
X^\mu
\end{array}
\right),
\label{erpa}
\end{eqnarray}
where
the matrix elements are given by
\begin{eqnarray}
A(\alpha\alpha':\lambda\lambda')&=&\langle0|[[:a^{\dag}_{\alpha'}a_\alpha:,\hat{H}],:a^{\dag}_\lambda a_{\lambda'}:]|0\rangle, \\
B(\alpha\beta\alpha'\beta':\lambda\lambda')
&=&\langle0|[[:a^{\dag}_{\alpha'}a^{\dag}_{\beta'}a_\beta a_\alpha:,\hat{H}],:a^{\dag}_\lambda a_{\lambda'}:]|0\rangle, \\
C(\alpha\alpha':\lambda_1\lambda_2\lambda_1'\lambda_2')
&=&\langle0|[[:a^{\dag}_{\alpha'}a_\alpha:,\hat{H}],
:a^{\dag}_{\lambda_1}a^{\dag}_{\lambda_2}a_{\lambda_2'}a_{\lambda_1'}:]|0\rangle, \\
D(\alpha\beta\alpha'\beta':\lambda_1\lambda_2\lambda_1'\lambda_2')
&=&\langle0|[[:a^{\dag}_{\alpha'}a^{\dag}_{\beta'}a_\beta a_\alpha:,\hat{H}],
:a^{\dag}_{\lambda_1}a^{\dag}_{\lambda_2}a_{\lambda_2'}a_{\lambda_1'}:]|0\rangle,
\label{D}\\
S_1(\alpha\alpha':\lambda\lambda')&=&\langle0|[:a^{\dag}_{\alpha'}a_\alpha:,:a^{\dag}_\lambda a_{\lambda'}:]|0\rangle, \\
T_1(\alpha\alpha':\lambda_1\lambda_2\lambda_1'\lambda_2')
&=&\langle0|[:a^{\dag}_{\alpha'}a_\alpha:,
:a^{\dag}_{\lambda_1}a^{\dag}_{\lambda_2}a_{\lambda_2'}a_{\lambda_1'}:]|0\rangle, \\
T_2(\alpha\beta\alpha'\beta':\lambda\lambda')
&=&\langle0|[:a^{\dag}_{\alpha'}a^{\dag}_{\beta'}a_\beta a_\alpha:,:a^{\dag}_\lambda a_{\lambda'}:]|0\rangle, \\
S_2(\alpha\beta\alpha'\beta':\lambda_1\lambda_2\lambda_1'\lambda_2')
&=&\langle0|[:a^{\dag}_{\alpha'}a^{\dag}_{\beta'}a_\beta a_\alpha:,
:a^{\dag}_{\lambda_1}a^{\dag}_{\lambda_2}a_{\lambda_2'}a_{\lambda_1'}:]|0\rangle.
\end{eqnarray}
At this point we want to insist on the fact that in eqs.(\ref{gs1})-(\ref{gs3}) 
antisymmetrisation is fully respected and, thus, the matrix elements entering 
eq. (\ref{erpa}) also respect antisymmetrisation fully. This is at variance of most 
other extensions of the RPA approach.

In the following we show that eq. (\ref{erpa}) can also be derived from the small amplitude limit of TDDM (eq. (\ref{stddm})).
We transform the eigenvector in eq. (\ref{stddm}) 
using an extended norm matrix including three-body components as
\begin{eqnarray}
\left(
\begin{array}{c}
x^\mu\\
X^\mu\\
Y^\mu
\end{array}
\right)=
\left(
\begin{array}{ccc}
S_1 &T_1 & T_{13} \\
T_2& S_2& T_{23} \\
T_{31} & T_{32} & T_{33}
\end{array}
\right)
\left(
\begin{array}{c}
x_\mu\\
y_\mu\\
z_\mu
\end{array}
\right),
\end{eqnarray}
where
\begin{eqnarray}
T_{31}&=&\langle0|[:a^{\dag}_{\alpha'}a^{\dag}_{\beta'}a^{\dag}_{\gamma'}a_\gamma a_\beta a_\alpha:,
:a^{\dag}_{\lambda}a_{\lambda'}:]|0\rangle,\\
T_{32}&=&\langle0|[:a^{\dag}_{\alpha'}a^{\dag}_{\beta'}a^{\dag}_{\gamma'}a_\gamma a_\beta a_\alpha:,
:a^{\dag}_{\lambda_1}a^{\dag}_{\lambda_2}a_{\lambda_2'}a_{\lambda_1'}:]|0\rangle\\
T_{13}&=&\langle0|[:a^{\dag}_{\alpha'} a_\alpha:,
:a^{\dag}_{\lambda_1}a^{\dag}_{\lambda_2}a^{\dag}_{\lambda_3}a_{\lambda_3'}a_{\lambda_2'}a_{\lambda_1'}:]|0\rangle,\\
T_{23}&=&\langle0|[:a^{\dag}_{\alpha'}a^{\dag}_{\beta'} a_\beta a_\alpha:,
:a^{\dag}_{\lambda_1}a^{\dag}_{\lambda_2}a^{\dag}_{\lambda_3}a_{\lambda_3'}a_{\lambda_2'}a_{\lambda_1'}:]|0\rangle,\\
T_{33}&=&\langle0|[:a^{\dag}_{\alpha'}a^{\dag}_{\beta'}a^{\dag}_{\gamma'}a_\gamma a_\beta a_\alpha:,
:a^{\dag}_{\lambda_1}a^{\dag}_{\lambda_2}a^{\dag}_{\lambda_3}a_{\lambda_3'}a_{\lambda_2'}a_{\lambda_1'}:]|0\rangle.
\end{eqnarray}
Then eq. (\ref{stddm}) is written as 
\begin{eqnarray}
\left(
\begin{array}{ccc}
aS_1+cT_2 &aT_1+cS_2 & aT_{13}+cT_{23} \\
bS_1+dT_2+eT_{31}& bT_1+dS_2+eT_{32}& bT_{13}+dT_{23}+eT_{33} \\
fS_{1}+gT_{2}+hT_{31} & fT_1+gS_2+hT_{32} & fT_{13}+gT_{23}+hT_{33}
\end{array}
\right)
\left(
\begin{array}{c}
x^\mu\\
X^\mu\\
Y^\mu
\end{array}
\right)=\Omega_\mu
\left(
\begin{array}{ccc}
S_1 &T_1 & T_{13} \\
T_2& S_2& T_{23} \\
T_{31} & T_{32} & T_{33}
\end{array}
\right)
\left(
\begin{array}{c}
x^\mu\\
X^\mu\\
Y^\mu
\end{array}
\right).
\label{stddm3}
\end{eqnarray}
The matrix elements on the left-hand side of the above equation can be written in the form of 
the double commutators with the hamiltonian \cite{ts08} except for those with $f$, $g$ and $h$:
Since $[:a^{\dag}_{\alpha'}a^{\dag}_{\beta'}a^{\dag}_{\gamma'}a_\gamma a_\beta a_\alpha:,\hat{H}]$ contains four-body operators,
the matrices $fS_{1}+gT_{2}+hT_{31}$, $fT_1+gS_2+hT_{32}$, and $fT_{13}+gT_{23}+hT_{33}$ cannot be expressed using the double
commutator with the hamiltonian. The matrices 
$A$, $B$, $C$, and $D$ in eq. (\ref{erpa}) have the following relations with $a$, $b$, $c$, $d$, and $e$ \cite{ts08},
$A=aS_1+cT_2$,
$B=bS_1+dT_2+eT_{31}$,
$C=aT_1+cS_2$, and
$D=bT_1+dS_2+eT_{32}$.
The expressions for $S_1$, $T_1$, $T_2$, $S_2$, $T_{31}$, and $T_{32}$ are given in ref. \cite{ts08}.
Thus it is shown that eq. (\ref{erpa}) is obtained from eq. (\ref{stddm3}) by neglecting the three-body sections.

As has been discussed in detail in ref. \cite{ts07}, the hamiltonian matrix on the left hand side of eq. (\ref{erpa}) 
is hermitian due to the ground-state conditions eqs. (\ref{gs1})-(\ref{gs3}) which guarantee the Jacobi's identity
\begin{eqnarray}
\langle 0|[[\hat{Q},\hat{H}],\hat{P}]|0\rangle -\langle 0|[[\hat{P},\hat{H}],\hat{Q}]|0\rangle=\langle 0|[\hat{H},[\hat{P},\hat{Q}]]|0\rangle=0,
\label{jacobi}
\end{eqnarray} 
where $\hat{P}$ and $\hat{Q}$ are either one-body or two-body operators. 
It is interesting to note that reversing the order of arguments and imposing hermiticity of the matrix on 
the left-hand side of eq.(\ref{erpa}) one can arrive at eqs. (\ref{gs1})-(\ref{gs3}).  
We remind in this context that Rowe \cite{Rowe} achieved hermiticity of his equation of motion method 
taking the arithmetic mean of the off diagonal elements. 
This somewhat artificial procedure finds a natural solution 
with the considerations given in our present procedure. 
It, therefore, can be stated that in the equation of motion method 
the non hermitian part of the matrices has to be put to zero what imposes the fulfillment of extra equations.

As mentioned above, the hamiltonian matrix of eq. (\ref{stddm3}) is not hermitian because of the asymmetry in the three-body parts:
It can be stated that eq. (\ref{erpa}) is obtained omitting the non-hermitian parts of eq. (\ref{stddm3}).
The commutator $[\hat{P},\hat{Q}]$ for three-body operators 
$\hat{P}$ and $\hat{Q}$ contains three-body, four-body and five-body
operators. In order to fulfill the condition eq. (\ref{jacobi}) and make an extended RPA 
with the three-body amplitude $Y^\mu$ hermitian, therefore, 
we need to consider two additional ground-state conditions for four-body and five-body operators. 
The problem of a further extended RPA with the three-body amplitude is beyond the scope of this paper.

The ortho-normal condition of ERPA (eq. (\ref{erpa})) is given as \cite{Taka}
\begin{eqnarray}
\left(
{x^\mu}^*~~{X^\mu}^*
\right)\left(
\begin{array}{cc}
S_1 & T_1 \\
T_2 & S_2
\end{array}
\right)\left(
\begin{array}{c}
x^{\mu'} \\
X^{\mu'}
\end{array}
\right)
=\pm\delta_{\mu\mu'},
\label{orthogonal}
\end{eqnarray}
where the negative sign is for a negative-energy solution. 
Equation (\ref{erpa}) is equivalent to the second RPA \cite{Saw,Wam}
when the ground-state correlations are neglected: Neglect of the ground-state correlations means that
$n_{\alpha\alpha'}=\delta_{\alpha\alpha'}(0)$ 
for occupied (unoccupied) states, $C_{\alpha\beta\alpha'\beta'}=0$ and $C_{\alpha\beta\gamma\alpha'\beta'\gamma}=0$.
In this limit, the one-body section of
eq. (\ref{erpa}), $Ax^\mu=\omega_\mu S_1x^\mu$, is equivalent to standard RPA.
We make a comparison with a simplified version of ERPA where the three-body correlation matrix $C_{\alpha\beta\gamma\alpha'\beta'\gamma'}$
is neglected in the calculation of the ground state. We refer this ERPA as to ERPA1. 
When eq. (\ref{erpa}) is solved numerically, first the norm matrix on the right-hand side of eq. (\ref{erpa}) is diagonalized.
Then eq. (\ref{erpa}) is solved in the truncated space consisting of the eigenstates of the norm matrix with nonvanishing eigenvalues.

\subsection{Stability condition}
We discuss the relation of the ground state conditions eqs. (\ref{gs1}) and (\ref{gs2}) to
the variational principle for the energy. We also discuss the stability 
condition of the ground state.
Suppose the following variational energy
\begin{eqnarray}
E=\langle 0|e^{-i\hat{F}}\hat{H}e^{i\hat{F}}|0\rangle,
\end{eqnarray}
where $\hat{F}$ is an arbitrary hermitian operator given as
\begin{eqnarray}
\hat{F}=\sum_{\alpha\alpha'}f_{\alpha\alpha'}:a^{\dag}_\alpha a_{\alpha'}:
+\sum_{\alpha\beta\alpha'\beta'}F_{\alpha\beta\alpha'\beta'}:a^{\dag}_{\alpha}a^{\dag}_{\beta}a_{\beta'}a_{\alpha'}:.
\label{variation0}
\end{eqnarray}
The energy $E$ can be expressed as
\begin{eqnarray}
E&=&\langle 0|\hat{H}|0\rangle+i\langle 0|[\hat{H},\hat{F}]|0\rangle+\frac{1}{2}\langle 0|[[\hat{F},\hat{H}],\hat{F}]|0\rangle+\cdot\cdot\cdot\cdot
\nonumber \\
&=&\langle 0|\hat{H}|0\rangle+i\sum_{\alpha\alpha'}\langle 0|[\hat{H},:a^{\dag}_\alpha a_{\alpha'}:]|0\rangle f_{\alpha\alpha'}
\nonumber \\
&+&i\sum_{\alpha\beta\alpha'\beta'}\langle 0|[\hat{H},:a^{\dag}_{\alpha}a^{\dag}_{\beta}a_{\beta'}a_{\alpha'}:]|0\rangle F_{\alpha\beta\alpha'\beta'}
\nonumber \\
&+&\frac{1}{2}\left(
{f}^*~~{F}^*
\right)\left(
\begin{array}{cc}
A & C \\
B & D
\end{array}
\right)\left(
\begin{array}{c}
f \\
F
\end{array}
\right)+\cdot\cdot\cdot\cdot.
\label{variation}
\end{eqnarray}
The stationary conditions eqs. (\ref{gs1}) and (\ref{gs2}) correspond to $\delta E/\delta f_{\alpha\alpha'}=0$ and $\delta E/\delta F_{\alpha\beta\alpha'\beta'}=0$,
respectively, and the single-particle hamiltonian of eq. (\ref{hfeq}) is obtained from 
the variation $\delta \langle 0|\hat{H}|0\rangle/\delta n_{\alpha\alpha'}=h_{\alpha\alpha'}$. The condition eq. (\ref{gs3}) 
can also be obtained by adding a three-body operator to eq. (\ref{variation0}).
If the total wavefunction $|0\rangle$ were used to calculate $n_{\alpha\alpha'}$,
 $C_{\alpha\beta\alpha'\beta'}$, $C_{\alpha\beta\gamma\alpha'\beta'\gamma'}$, and 
also the four-body correlation matrix,
eqs.(\ref{gs1})-(\ref{gs3}) would correspond to the variational principle
for the total wavefuntion. However, it is impracticable to find $|0\rangle$
itself. In TDDM, eqs.(\ref{gs1})-(\ref{gs3}) are used to obtain not the total wavefunction but 
$n_{\alpha\alpha'}$, $C_{\alpha\beta\alpha'\beta'}$, and $C_{\alpha\beta\gamma\alpha'\beta'\gamma'}$ 
by approximating a four-body density matrix with lower level density matrices. In this sense, TDDM deviates from
the variational principle for the total wavefunction, although
eqs.(\ref{gs1})-(\ref{gs3}) can be formally derived from the variational principle.
 
The stability matrix in the last line on the right-hand
side of eq. (\ref{variation}) is  
nothing but the hamiltonian matrix of eq. (\ref{erpa}). As was mentioned 
above, the condition for the three-body 
matrix eq. (\ref{gs3}) in addition to eqs. (\ref{gs1}) and (\ref{gs2}) 
is necessary to make the stability matrix hermitian although
the operators eqs. (\ref{exoper}) and (\ref{variation0}) are
at most two-body ones.
The stability condition that $|0\rangle$ corresponds to a minimum 
in the energy surface is given by
\begin{eqnarray} 
\left(
{f}^*~~{F}^*
\right)\left(
\begin{array}{cc}
A & C \\
B & D
\end{array}
\right)\left(
\begin{array}{c}
f \\
F
\end{array}
\right)\ge 0
\label{stability}
\end{eqnarray}
for arbitrary $f_{\alpha\alpha'}$ and $F_{\alpha\beta\alpha'\beta'}$. This condition is satisfied when the stability matrix
is positive definite. In this case, eq. (\ref{erpa}) has only real eigenvalues since the norm matrix 
in eq. (\ref{erpa}) is hermitian \cite{RS}.
\subsection{Relation to self-consistent RPA}
In this subsection we discuss some relation of the one-body part of eq. (\ref{erpa}) 
\begin{eqnarray}
Ax^\mu=\omega_\mu S_1x^\mu
\label{SCRPA}
\end{eqnarray}
to SCRPA [14].
The matrix $A$
is explicitly written as
\begin{eqnarray}
A(\alpha\alpha':\lambda\lambda')&=&
(\epsilon_\alpha-\epsilon_{\alpha'})
(n_{\lambda'\alpha'}\delta_{\alpha\lambda}
-n_{\alpha\lambda}\delta_{\alpha'\lambda'})
\nonumber \\
&+&\sum_{\gamma\gamma'}[(\langle\alpha\gamma|v|\gamma'\lambda\rangle_A
n_{\lambda'\gamma}
-\langle\alpha\lambda'|v|\gamma'\gamma\rangle_A
n_{\gamma\lambda})n_{\gamma'\alpha'}
\nonumber \\
&-&(\langle\gamma'\gamma|v|\alpha'\lambda\rangle_A
n_{\lambda'\gamma}
-\langle\gamma'\lambda'|v|\alpha'\gamma\rangle_A
n_{\gamma\lambda})n_{\alpha\gamma'}]
\nonumber \\
&-&\sum_{\gamma\gamma'\gamma''}(\langle\alpha\gamma|v|\gamma'\gamma''\rangle
C_{\gamma'\gamma''\lambda\gamma}\delta_{\alpha'\lambda'}
+\langle\gamma\gamma'|v|\alpha'\gamma''\rangle
C_{\lambda'\gamma''\gamma\gamma'}\delta_{\alpha\lambda})
\nonumber \\
&+&\sum_{\gamma\gamma'}(\langle\alpha\gamma|v|\lambda\gamma'\rangle_A
C_{\lambda'\gamma'\alpha'\gamma}
+\langle\lambda'\gamma|v|\alpha'\gamma'\rangle_A
C_{\alpha\gamma'\lambda\gamma})
\nonumber \\
&-&\sum_{\gamma\gamma'}(\langle\alpha\lambda'|v|\gamma\gamma'\rangle
C_{\gamma\gamma'\alpha'\lambda}
+\langle\gamma\gamma'|v|\alpha'\lambda\rangle
C_{\alpha\lambda'\gamma\gamma'}),
\label{A-term}
\end{eqnarray}
where the subscript $A$ means that the corresponding matrix is antisymmetrized.
The norm matrix $S_1$ is given by
\begin{eqnarray}
S_1(\alpha\alpha':\lambda\lambda')=n_{\lambda'\alpha'}\delta_{\alpha\lambda}
-n_{\alpha\lambda}\delta_{\alpha'\lambda'}.
\end{eqnarray}
The first two terms with $C_{\alpha\beta\alpha'\beta'}$ 
on the right-hand side of eq. (\ref{A-term}) describe the 
self-energy of the particle - hole state due to ground-state correlations \cite{Janssen}.
The last four terms with $C_{\alpha\beta\alpha'\beta'}$ may be
interpreted as the modification of the particle - hole interaction
caused by ground-state correlations \cite{Janssen}.
Equation (\ref{SCRPA}) is formally the same as the SCRPA equation. Both equations include
the effects of ground-state correlations through 
$n_{\alpha\alpha'}$ and $C_{\alpha\beta\alpha'\beta'}$.
The difference between eq. (\ref{SCRPA}) and the SCRPA equation
lies in the fact that $n_{\alpha\alpha'}$ and $C_{\alpha\beta\alpha'\beta'}$
are self-consistently generated from $x^\mu_{\alpha\alpha'}$ in SCRPA, while they are given by eqs. (\ref{gs1}) and (\ref{gs2}), 
independently of eq. (\ref{SCRPA}).
Another difference stems from the fact that eq. (\ref{gs1}) is used in SCRPA 
to determine the optimal single particle basis whereas the occupation numbers 
are obtained from a separate relation \cite{Dukelsky}.

\section{Application to Lipkin model\label{sect.3}}
\subsection{Lipkin model}

The Lipkin model \cite{Lip} describes an N-fermions system with two
N-fold degenerate levels with energies $\epsilon/2$ and $-\epsilon/2$,
respectively. The upper and lower levels are labeled by quantum number
$p$ and $-p$, respectively, with $p=1,2,...,N$. We consider
the standard hamiltonian
\begin{equation}
\hat{H}=\epsilon \hat{J}_{z}+\frac{V}{2}(\hat{J}_+^2+\hat{J}_-^2),
\label{elipkin}
\end{equation}
where the operators are given as
\begin{eqnarray}
\hat{J}_z=\frac{1}{2}\sum_{p=1}^N(a_p^{\dag}a_p-{a_{-p}}^{\dag}a_{-p}), \\
\hat{J}_{+}=\hat{J}_{-}^{\dag}=\sum_{p=1}^N a_p^{\dag}a_{-p}.
\end{eqnarray}

Since eq. (\ref{gs1}) 
keeps the property $n_{p-p}=n_{-pp}=0$, the single-particle state
in eq. (\ref{hfeq}) is equivalent to the original single-particle basis labeled by $-p$ and $p$. 
When the matrix inversion in eq. (\ref{gradient}) is taken,
it is essential to preserve symmetry properties of all the matrices.
Therefore, we use the so-called m scheme to define each matrix throughout our numerical calculations:
For example, $n_{-p-p}$, $n_{-p'-p'}$, $C_{-p-p'pp'}$, $C_{-p'-ppp'}$, 
$C_{-p-p'p'p}$ and $C_{-p'-pp'p}$  with $p$ and $p'=1,2,...,N$ 
are treated as independent quantities. Of course, eqs. (\ref{gs1}) and (\ref{gs2}) guarantee
$n_{-p-p}$=$n_{-p'-p'}$ and $C_{-p-p'pp'}=-C_{-p'-ppp'}=-C_{-p-p'p'p}=C_{-p'-pp'p}$.
Let us mention that due to the use of the m-scheme 
the dimension of the matrices can become considerable, 
even for the case of the Lipkin model. One is forced to do that, 
otherwise the full antisymmetry can not be maintained. 

\subsection{Ground state}

Since eq. (\ref{gradient}) involves the time-consuming inversion of a large matrix, 
we first find an approximate solution for eqs. (\ref{gs1}) - (\ref{gs3})
to be used in the gradient method using the following time-dependent approach \cite{TTS,pfitz}:
Starting with the HF ground state, we solve eqs. (\ref{td1})-(\ref{td3})
using the time-dependent interaction $V(t)=V\times t/\tau$ with large $\tau$ \cite{pfitz,t09}.
We tested the reliability of this time-dependent method for an $N=2$ system, 
to which the truncation of the reduced density matrices up to the two-body level should give the exact solution. 
Since there are no three-body density matrices, Eq. (\ref{td2}) for the
$N=2$ system is modified to
\begin{eqnarray}
i\hbar\frac{d}{dt}\rho_{\alpha\beta\alpha'\beta'}&=&
(\epsilon_{\alpha}+\epsilon_{\beta}-\epsilon_{\alpha'}-\epsilon_{\beta'})
\rho_{\alpha\beta\alpha'\beta'}
\nonumber \\
&+&\sum_{\lambda_1\lambda_2}[
\langle\alpha\beta|v|\lambda_1\lambda_2\rangle\rho_{\lambda_1\lambda_2\alpha'\beta'}
-\langle\lambda_1\lambda_2|v|\alpha'\beta'\rangle\rho_{\alpha\beta\lambda_1\lambda_2}],
\end{eqnarray}
where $\rho_{\alpha\beta\alpha'\beta'}={\cal A}(n_{\alpha\alpha'}n_{\beta\beta'})+C_{\alpha\beta\alpha'\beta'}$.
We found that $\tau=2\pi\hbar/\epsilon \times 10$ is sufficiently large to suppress the spurious mixing of
excited states and the obtained ground-state energy is equivalent to the exact one $-\sqrt{\epsilon^2+V^2}$
within numerical accuracy.
After the time-dependent calculation
eq. (\ref{gradient}) is then solved for a few hundred iterations to achieve the conditions
eqs. (\ref{gs1}) - (\ref{gs3}) for each matrix element,
which guarantee the hermiticity of the hamiltonian matrix of eq. (\ref{erpa}). 
Since the time dependent approach already gives a good approximate
solution for eqs. (\ref{gs1}) - (\ref{gs3}), the change in the total energy during the iteration process
of eq. (\ref{gradient}) is negligible.
\begin{figure}
\begin{center}
\includegraphics[height=6cm]{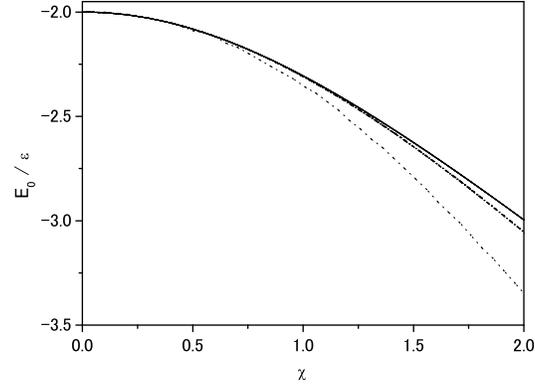}
\end{center}
\caption{Ground-state energy  in TDDM (solid line) as a function of $\chi=(N-1)|V|/\epsilon$ for $N=4$. 
The dotted line depicts the results in TDDM1 where the three-body correlation matrix is neglected, 
and the dot-dashed line the exact values. }
\label{fig1}
\end{figure}
\begin{figure}
\begin{center}
\includegraphics[height=6cm]{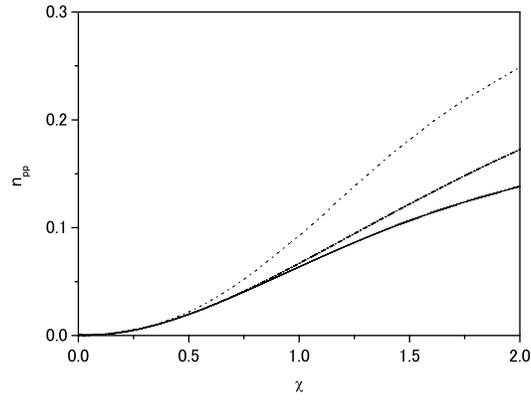}
\end{center}
\caption{Occupation probability of the upper state as a function of $\chi$ for $N=4$. The meaning of the three
lines is the same as in fig. \ref{fig1}.}
\label{fig2}
\end{figure}
\begin{figure}
\begin{center}
\includegraphics[height=6cm]{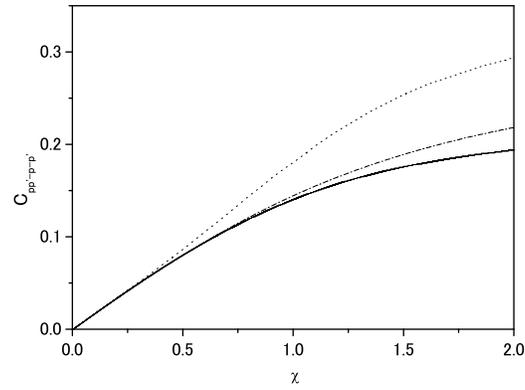}
\end{center}
\caption{Two-body correlation matrix $C_{pp'-p-p'}$ as a function of $\chi$ for $N=4$.
The meaning of the three
lines is the same as in fig. \ref{fig1}.}
\label{fig3}
\end{figure}
\begin{figure}
\begin{center}
\includegraphics[height=6cm]{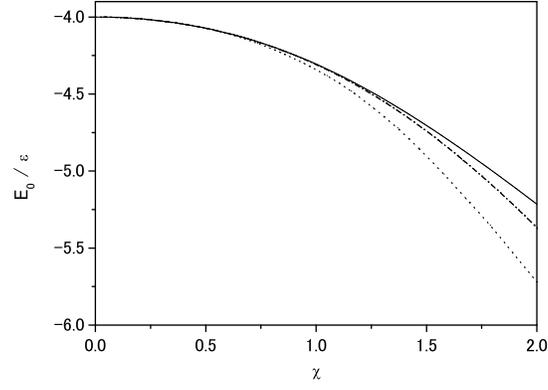}
\end{center}
\caption{Ground-state energy as a function of $\chi$ for $N=8$.
The meaning of the three
lines is the same as in fig. \ref{fig1}.}
\label{fig4}
\end{figure}

\begin{figure}
\begin{center}
\includegraphics[height=6cm]{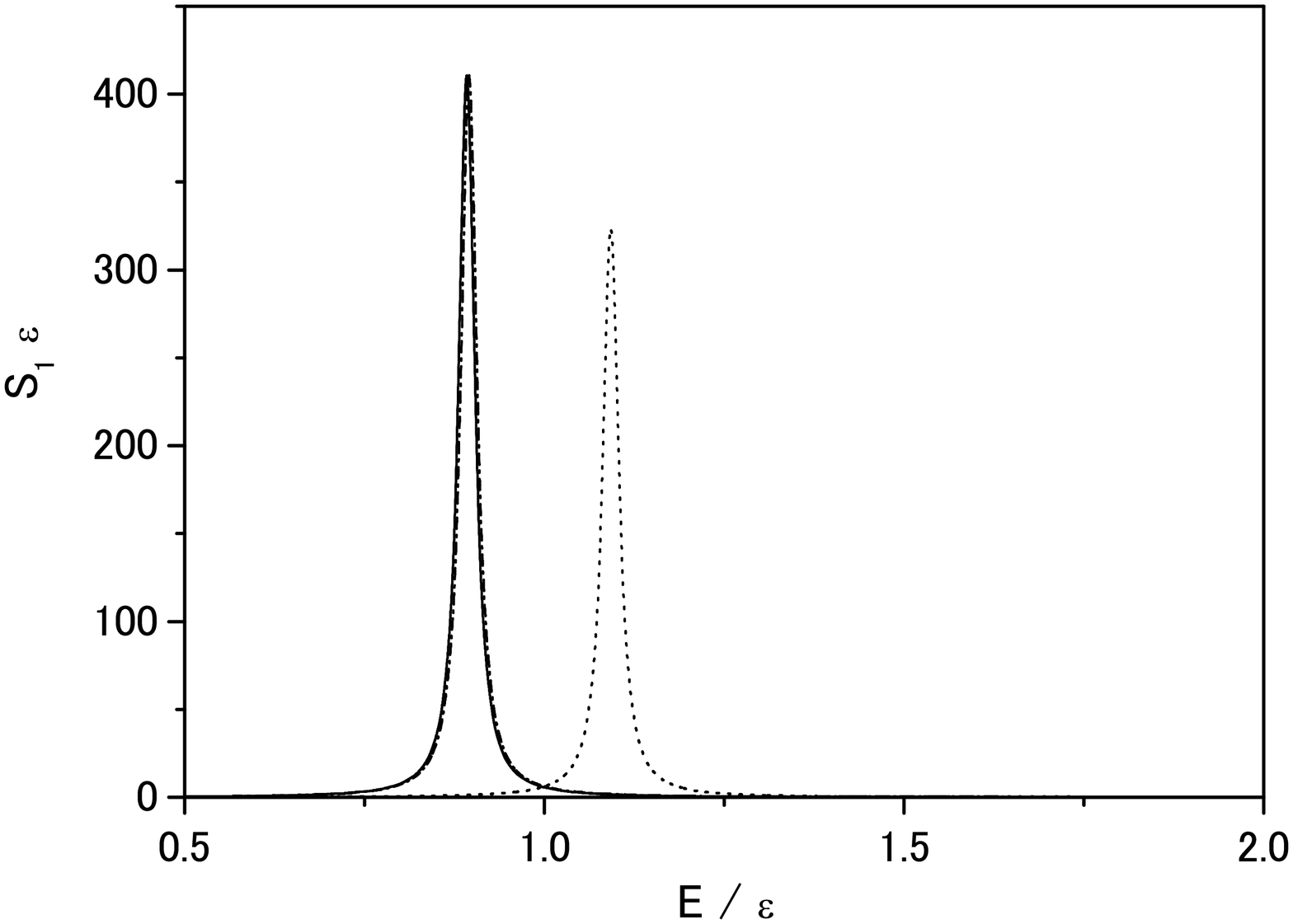}
\end{center}
\caption{Strength function of the one-phonon state in ERPA (solid line) for $\chi=1$ and $N=4$.
The dotted and dot-dashed lines depict the result in ERPA1 and the exact solution, respectively.}
\label{fig5}
\end{figure}
  
\begin{figure}
\begin{center}
\includegraphics[height=6cm]{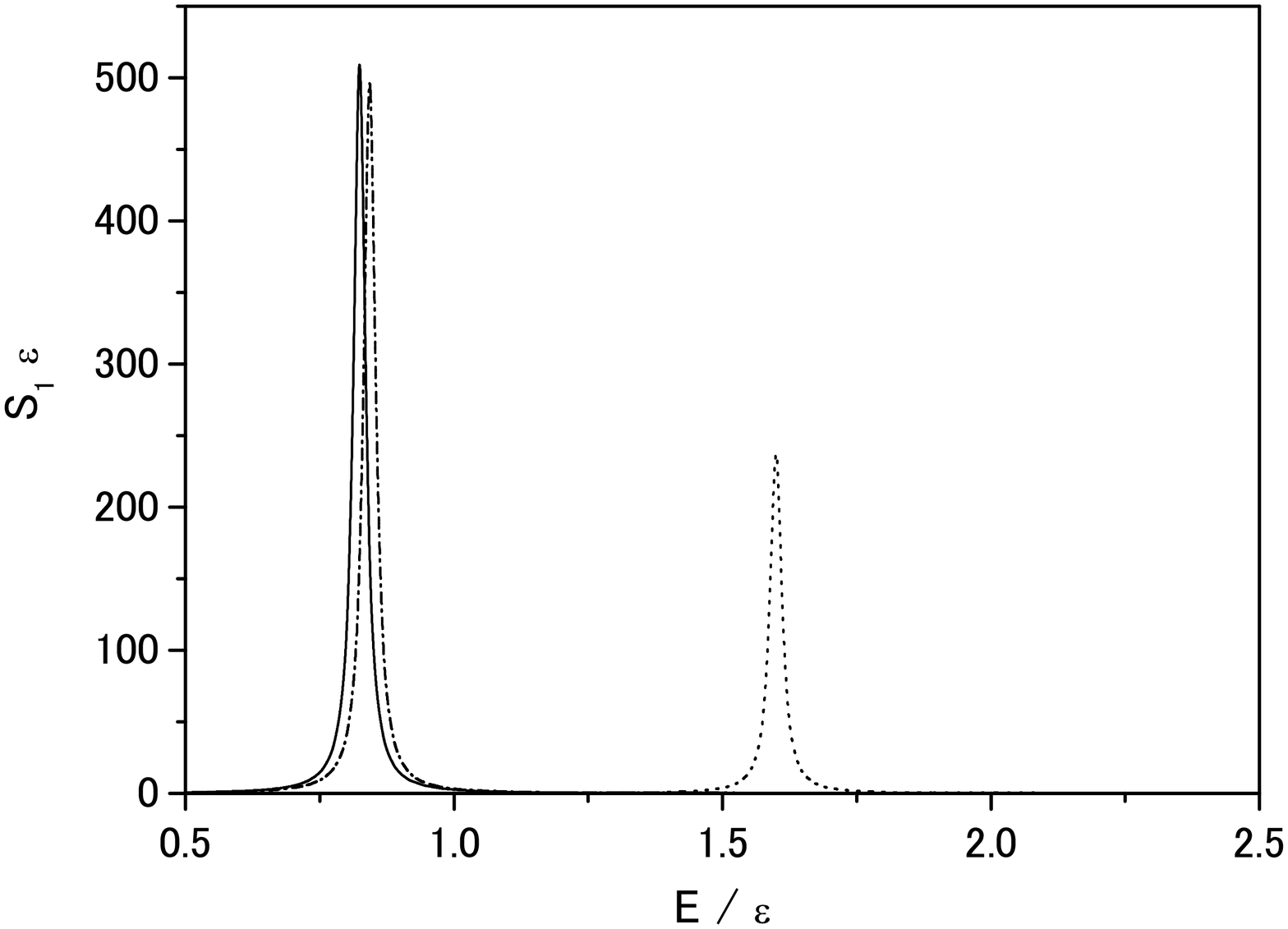}
\end{center}
\caption{Strength function of the one-phonon state for $\chi=1.5$ and $N=4$.
The meaning of the three
lines is the same as in fig. \ref{fig5}.}
\label{fig7}
\end{figure}

\begin{figure}
\begin{center}
\includegraphics[height=6cm]{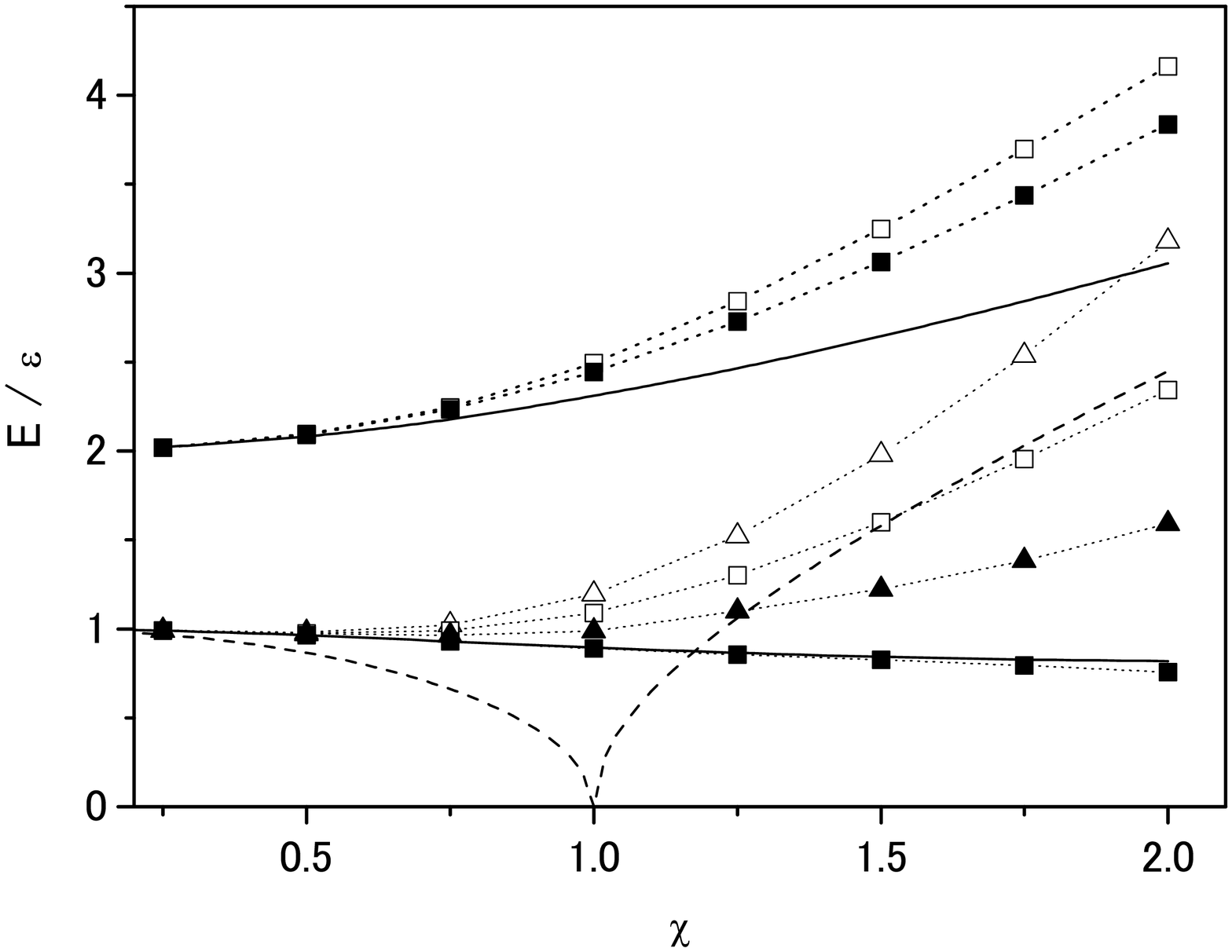}
\end{center}
\caption{Excitation energies of the one-phonon state (lower part) and the two-phonon state (upper part) 
as a function of $\chi$ for $N=4$.
The filled and open squares depict the results in ERPA and ERPA1, respectively.
The filled and open triangles indicate the results in IRPA and IRPA1, respectively.
The dashed line depicts the results in RPA. 
The exact solutions are given by the solid lines.} 
\label{fig9}
\end{figure}

The ground-state energy in TDDM
(solid line) is shown in fig. \ref{fig1} as a function of $\chi=(N-1)V/\epsilon$ for $N=4$.
The results with TDDM1
and the exact ground-state energies 
are shown in fig. \ref{fig1} with the dotted and dot-dashed lines, respectively. 
As seen in fig. \ref{fig1}, the ground-state energies in TDDM1 are overestimated.
This unpleasant feature of TDDM1 
is removed by the inclusion of the three-body correlation matrix
and the agreement with the exact solution is much improved.
In order to see the ground-state properties in more detail, we show 
the occupation probability of the upper level $n_{pp}$ and the two-body correlation matrix 
$C_{pp'-p-p'}$ in figs. \ref{fig2} and \ref{fig3}, respectively. The solid and dotted lines denote the 
results in TDDM and TDDM1, respectively. The exact values are shown with the dot-dashed lines.
Figures  \ref{fig2} and \ref{fig3} show that
the agreement of these matrices with the exact ones is also improved 
by the inclusion of the three-body correlation matrices. 
As is understood from eqs. (\ref{F2}) and (\ref{T}) in the Appendix, the three-body correlation matrix
interferes in particle - particle and particle - hole correlations
and presumably plays a role in screening these two-body correlations.
In order to investigate the effects of the three-body correlations in larger $N$ systems,
we calculate the ground-state energies in TDDM for $N=8$. As shown in fig. \ref{fig4}, an improvement similar to the $N=4$ case is achieved.

The importance of the three-body correlation matrix 
in the standard Lipkin model is
strongly in contrast to the case of an extended Lipkin-model hamiltonian 
which we have previously studied \cite{mt07,ts08}.
The extended Lipkin model has the additional particle scattering term ($U$ term) \cite{Yang,Taka03}
\begin{eqnarray}
\frac{1}{2}U[\hat{J}_z(\hat{J}_++\hat{J}_-)+(\hat{J}_++\hat{J}_-)\hat{J}_z].
\end{eqnarray}
This $U$ term generates a mean-field potential \cite{Taka03} and gives nonzero off-diagonal
elements of the occupation matrix, that is, $n_{p-p}\neq0$ and $n_{-pp}\neq0$. 
Therefore, a significant portion of the 
interaction energy including the $V$ term in eq. (\ref{elipkin}) 
is carried as the mean-field energy \cite{mt07}, almost independently of the 
strength of the $U$ term. As a consequence, the two-body correlation matrix and the correlation energy become quite small
in the extended Lipkin model. 
The effect of the three-body correlations on the ground-state energy was 
found even smaller when the $U$ term was included \cite{ts08}.
In the case of the extended Lipkin model the ground-state energies calculated in TDDM1 are always above 
the exact values and such an 
overbound problem as in the standard Lipkin model does not occur.
Now the importance of the three-body ground-state correlations in the standard Lipkin model is 
understood as a consequence of lack of the mean-field energy, which makes the truncation of the reduced
density matrices up to the two-body level unreliable and enhances the relative importance of the three-body 
correlations.

The standard Lipkin model has {\it deformed} HF solutions with $n_{-pp}\neq 0$ for $\chi>1$ \cite{RS}, which consequently
carry the mean-field energy which sums up already a lot of correlations. 
One may then think that the truncation up to the two-body level be reliable when 
the two-body correlation matrix is defined using
the deformed HF single-particle states. To answer this question, we performed a simplified gradient-method calculation
where the three-body parts in eq. (\ref{gradient}) are neglected and a deformed HF state at $\chi=1.5$ is used as the initial state.
We found that the converged results are the same as those shown in figs. \ref{fig1}-\ref{fig3}: The residual interaction
plays a role in restoring the symmetry of eq. (\ref{elipkin}) which is violated by the deformed HF solution. 
In other words starting with a 'deformed' solution, at the end of the iteration cycle the solution is back to 'sphericity'.
Thus the use of the deformed HF basis does not improve the two-body level 
approximation for the standard Lipkin model. This somewhat surprising feature may be due to the small number of particles considered. 
It is known that standard deformed RPA becomes exact in the macroscopic limit. 
Therefore, we suppose that considering higher particle numbers, the deformation will catch on
when we start with the deformed HF state.

\subsection{Excited states}

For the evaluation of the excited states, we use the ERPA equation (25).
First we present the results for the one-phonon states in the $N=4$ system
excited by the operator $\hat{Q}=\hat{J}_++\hat{J}_-$.
Figures \ref{fig5} and \ref{fig7} show the strength functions of the one-phonon states at $\chi=1$ and $\chi=1.5$, respectively.
To facilitate easy comparison among various calculations, 
we smooth
the strength functions with an artificial width $\Gamma_{\rm FWHM}/\epsilon=0.025$.
The solid and dotted lines denote the results in ERPA and ERPA1, respectively. The exact results are shown with the
dot-dashed lines. 
The results in ERPA almost coincide with the exact ones, while those in ERPA1 are located above the
exact solutions. Note that RPA breaks down at $\chi=1$ \cite{RS} as shown in fig. \ref{fig9}. 
At $\chi=1.5$ the effect of the three-body correlations on the one-phonon state
is drastic because it significantly improves the occupation probabilities and the two-body correlation matrices
as shown in figs \ref{fig2} and \ref{fig3}. 
Thus it is found that the inclusion of
the three-body ground-state correlations also gives a better description of the one-phonon state in the standard Lipkin model.
The $\chi$ dependence of the excitation energy of the one-phonon state is shown in fig. \ref{fig9} 
for various approximations. The filled triangles indicate the results obtained from eq. (\ref{SCRPA})
using the same $n_{\alpha\alpha'}$ and $C_{\alpha\beta\alpha'\beta'}$ as those used in ERPA. This approximation is referred to as 
the improved RPA (IRPA). Similarly, the results obtained from eq. (\ref{SCRPA}) with 
the same $n_{\alpha\alpha'}$ and $C_{\alpha\beta\alpha'\beta'}$ as those used in ERPA1 are referred to as the IRPA1 results (open triangles).
The excitation energies of the one-phonon state in IRPA and IRPA1 are larger than the exact values and increase with increasing $\chi$.
This is explained by the increase in the self-energy terms in eq. (\ref{A-term}). Since the deviation of 
the occupation matrix and correlation matrix from the exact values is larger in TDDM1 than in TDDM, 
the excitation energies in IRPA1
are larger than those in IRPA.
The difference between the results in IRPA and ERPA and also between IRPA1 and ERPA1 is due to the coupling
of the one-body amplitudes to the two-body ones. 
Since the parity of the number of particle - hole pairs is conserved in the standard Lipkin model,
the one-body amplitudes can couple to the two-body amplitudes 
that have the same parity: For example, $x^\mu_{p-p}$ couples to $X^\mu_{pp'-pp'}$, which expresses excitations built 
on the two particle - two hole
configurations in the ground state.
As shown in fig. \ref{fig9}, the coupling to the two-body amplitudes is essential to obtain
the one-phonon states with accurate excitation energies.

\begin{figure}
\begin{center}
\includegraphics[height=6cm]{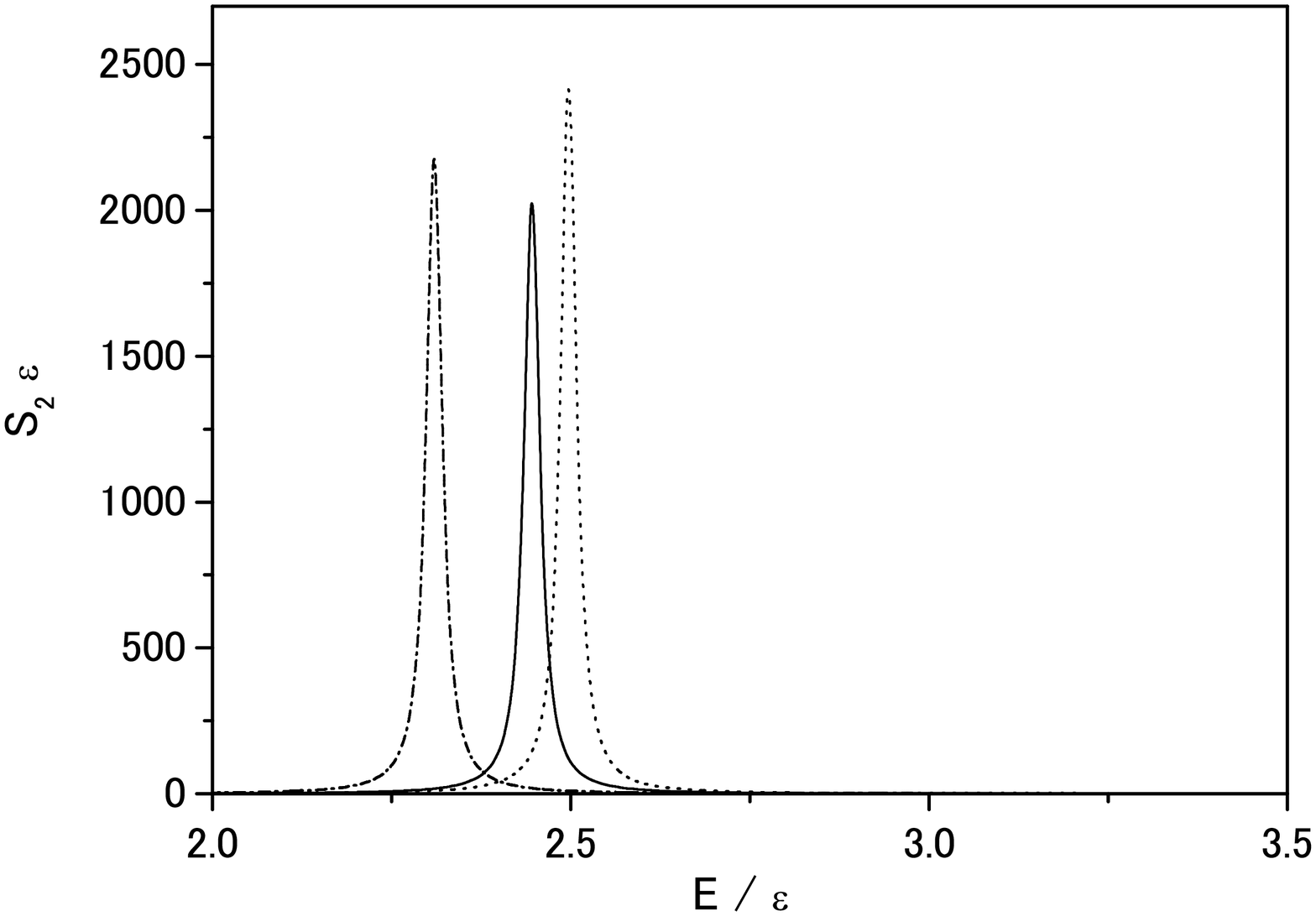}
\end{center}
\caption{Strength function of the two-phonon state in ERPA (solid line) for $\chi=1$ and $N=4$.
The dotted and dot-dashed lines depict the result in ERPA1 and the exact solution, respectively.}
\label{fig6}
\end{figure}

\begin{figure}
\begin{center}
\includegraphics[height=6cm]{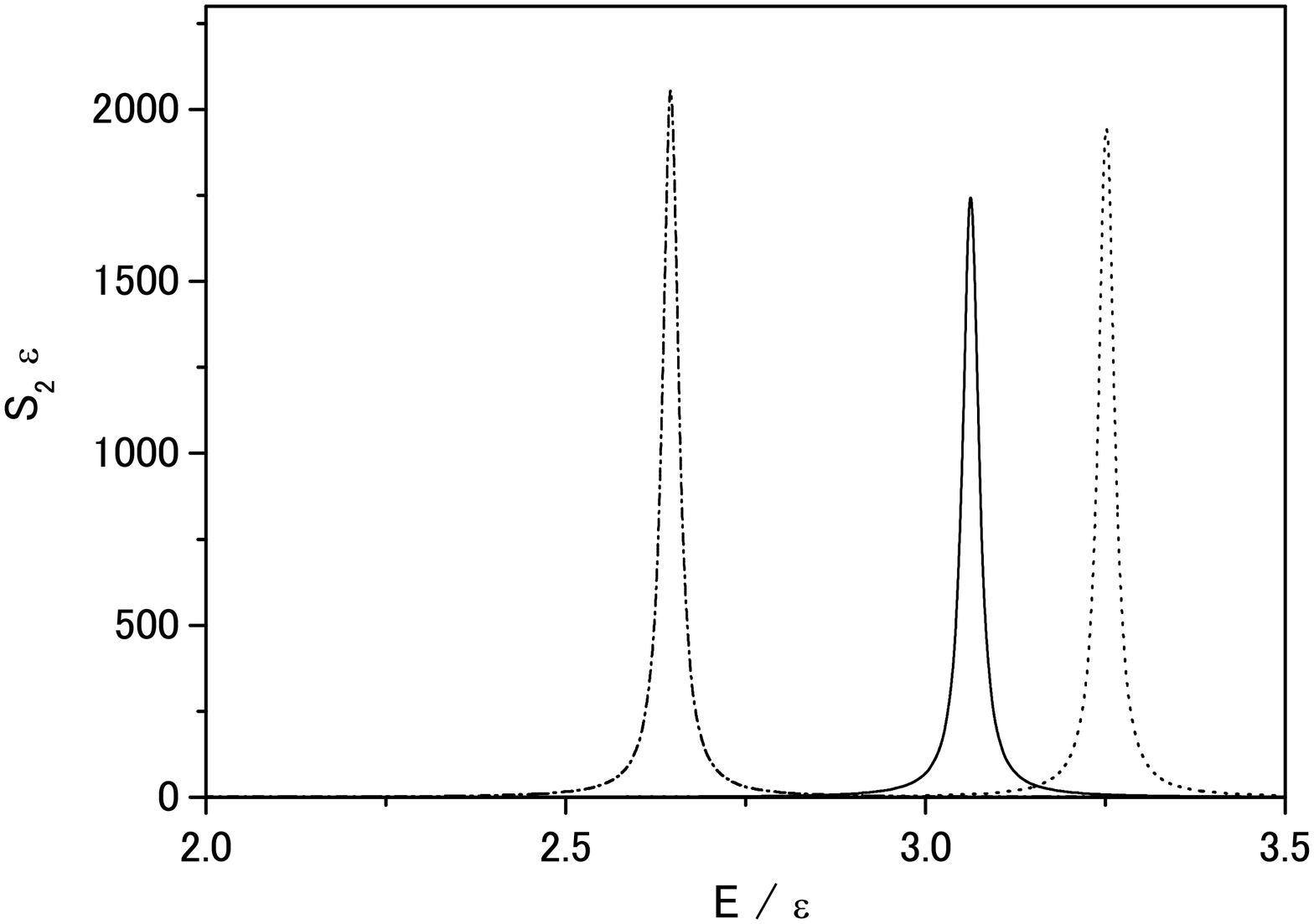}
\end{center}
\caption{Strength function of the two-phonon state for $\chi=1.5$ and $N=4$.
The meaning of the three
lines is the same as in fig. \ref{fig6}.}
\label{fig8}
\end{figure}

Now we discuss the two-phonon states excited by $\hat{Q}^2=(\hat{J}_++\hat{J}_-)^2$. The results at $\chi=1$ and 1.5 
are  shown in figs. \ref{fig6} and \ref{fig8}, respectively. Due to the parity conservation, the two-phonon state excited by 
$\hat{Q}^2$ does not couple to the one-phonon state in the standard Lipkin model. 
The excitation energies both in ERPA and ERPA1 are higher than the exact value and 
the improvement of the two-phonon state due to the inclusion of the three-body correlations 
is not so large as in the case of the one-phonon state. 
This indicates that the correlations among the two-body amplitudes are not sufficient to lower the
energy of the two-phonon state, suggesting the importance of the coupling to the three-body amplitudes
which are neglected in ERPA and ERPA1. The $\chi$ dependence of the excitation energy of the two-phonon state 
is also shown in fig. \ref{fig9}.

We also studied the stability condition by diagonalizing the stability matrix
eq. (\ref{stability}). We found that all the eigenvalues associated with the 
physical operators such as $J_+$, $J_-$ and $J_z$ which consist of the hamiltonian eq. (\ref{elipkin}) 
are positive definite. This corresponds to the real eigenvalues of eq. (\ref{erpa}) for the one-phonon and two-phonon states 
excited by $Q=J_++J_-$ and $Q^2$. However, the stability matrix has 
some negative eigenvalues for other degrees of freedom, though their absolute values are small; They are at most 1\% 
of the maximum positive eigenvalue for the physical operators. 
This means that the stability condition is not always satisfied for unphysical states. These states correspond to non-collective 
phonons as discussed in ref. \cite{Janssen}.

\section{Summary \label{sect.4}}
The density-matrix formalism which includes the effects of three-body correlations on the ground state 
was applied to the standard Lipkin model.
It was found that the inclusion of the three-body correlations removes 
unpleasant
features of the two-body level approximation and drastically improves the 
ground-state properties.
It was discussed that the importance of the three-body correlations in the standard Lipkin model is
attributed to the fact that it does not have the mean-field contributions
when the conservation of the number of particle - hole pairs is respected. 
The extended RPA built on the ground state with three-body correlations was applied to the one-phonon and 
two-phonon states. It was found that the spectrum of the one-phonon state is drastically 
improved by the inclusion of the three-body
correlations. In the case of the two-phonon states, however, the agreement with the exact solutions is not so good
as in the case of the one-phonon states, which suggests the importance of the coupling to the three-body amplitudes.
Inclusion of the three-body correlations for realistic nuclei may be impracticable.
Fortunately, mean-field effects are quite important in nuclei and the mean-field theories give good first description
for ground states and collective excitations. Therefore, the truncation
up to the two-body level may be justified as in the case of the extended Lipkin model. 
When the three-body correlation matrix is neglected, the hermiticity of the 
hamiltonian matrix in ERPA is not guaranteed. 
Our applications so far performed for realistic cases indicate that this does not cause any serious problems.

\appendix
\section{}
$F_1$, $F_2$ and $F_3$ in eqs. (\ref{gs1})-(\ref{gs3}) are shown.
The single-particle states which satisfy the HF-like equation $h\phi_\alpha=\epsilon\phi_\alpha$ are used.
\begin{eqnarray}
F_1(\alpha\alpha')&=&(\epsilon_{\alpha}-\epsilon_{\alpha'})n_{\alpha\alpha'}+
\sum_{\lambda_1\lambda_2\lambda_3}(C_{\lambda_1\lambda_2\alpha'\lambda_3}
\langle\alpha\lambda_3|v|\lambda_1\lambda_2\rangle -
C_{\alpha\lambda_3\lambda_1\lambda_2}
\langle\lambda_1\lambda_2|v|\alpha'\lambda_3\rangle),\\
F_2(\alpha\beta\alpha'\beta')&=&
(\epsilon_{\alpha}+\epsilon_{\beta}-\epsilon_{\alpha'}-\epsilon_{\beta'})
C_{\alpha\beta\alpha'\beta'}+
B_{\alpha\beta\alpha'\beta'}+P_{\alpha\beta\alpha'\beta'}+H_{\alpha\beta\alpha'\beta'}
+T_{\alpha\beta\alpha'\beta'}
\label{F2},
\\
F_3(\alpha\beta\gamma\alpha'\beta'\gamma')&=&(\epsilon_\alpha+\epsilon_\beta+\epsilon_\gamma
-\epsilon_{\alpha'}-\epsilon_{\beta'}-\epsilon_{\gamma'})
C_{\alpha\beta\gamma\alpha'\beta'\gamma'}
\nonumber \\
&+&I(\alpha\beta\gamma\alpha'\beta'\gamma')
-I(\beta\alpha\gamma\alpha'\beta'\gamma')
-I(\gamma\beta\alpha\alpha'\beta'\gamma')
\nonumber \\
&-&I^*(\alpha'\beta'\gamma'\alpha\beta\gamma)
+I^*(\beta'\alpha'\gamma'\alpha\beta\gamma)
+I^*(\gamma'\beta'\alpha'\alpha\beta\gamma)
\nonumber \\
&+&J(\alpha\beta\gamma\alpha'\beta'\gamma')
+J(\beta\gamma\alpha\alpha'\beta'\gamma')
-J(\alpha\gamma\beta\alpha'\beta'\gamma')
\nonumber \\
&-&J^*(\alpha'\beta'\gamma'\alpha\beta\gamma)
-J^*(\beta'\gamma'\alpha'\alpha\beta\gamma)
+J^*(\alpha'\gamma'\beta'\alpha\beta\gamma),
\end{eqnarray}
where
\begin{eqnarray}
B_{\alpha\beta\alpha'\beta'}&=&\sum_{\lambda_1\lambda_2\lambda_3\lambda_4}
\langle\lambda_1\lambda_2|v|\lambda_3\lambda_4\rangle_A
[(\delta_{\alpha\lambda_1}-n_{\alpha\lambda_1})(\delta_{\beta\lambda_2}-n_{\beta\lambda_2})
n_{\lambda_3\alpha'}n_{\lambda_4\beta'}
\nonumber \\
&-&n_{\alpha\lambda_1}n_{\beta\lambda_2}(\delta_{\lambda_3\alpha'}-n_{\lambda_3\alpha'})
(\delta_{\lambda_4\beta'}-n_{\lambda_4\beta'})],
\\
P_{\alpha\beta\alpha'\beta'}&=&\sum_{\lambda_1\lambda_2\lambda_3\lambda_4}
\langle\lambda_1\lambda_2|v|\lambda_3\lambda_4\rangle
[(\delta_{\alpha\lambda_1}\delta_{\beta\lambda_2}
-\delta_{\alpha\lambda_1}n_{\beta\lambda_2}
-n_{\alpha\lambda_1}\delta_{\beta\lambda_2})
C_{\lambda_3\lambda_4\alpha'\beta'}
\nonumber \\
&-&(\delta_{\lambda_3\alpha'}\delta_{\lambda_4\beta'}
-\delta_{\lambda_3\alpha'}n_{\lambda_4\beta'}
-n_{\lambda_3\alpha'}\delta_{\lambda_4\beta'})
C_{\alpha\beta\lambda_1\lambda_2}],
\\
H_{\alpha\beta\alpha'\beta'}&=&\sum_{\lambda_1\lambda_2\lambda_3\lambda_4}
\langle\lambda_1\lambda_2|v|\lambda_3\lambda_4\rangle_A
[\delta_{\alpha\lambda_1}(n_{\lambda_3\alpha'}C_{\lambda_4\beta\lambda_2\beta'}
-n_{\lambda_3\beta'}C_{\lambda_4\beta\lambda_2\alpha'})
\nonumber \\
&+&\delta_{\beta\lambda_2}(n_{\lambda_4\beta'}C_{\lambda_3\alpha\lambda_1\alpha'}
-n_{\lambda_4\alpha'}C_{\lambda_3\alpha\lambda_2\beta'})
\nonumber \\
&-&\delta_{\alpha'\lambda_3}(n_{\alpha\lambda_1}C_{\lambda_4\beta\lambda_2\beta'}
-n_{\beta\lambda_1}C_{\lambda_4\alpha\lambda_2\beta'})
\nonumber \\
&-&\delta_{\beta'\lambda_4}(n_{\beta\lambda_2}C_{\lambda_3\alpha\lambda_1\alpha'}
-n_{\alpha\lambda_2}C_{\lambda_3\beta\lambda_1\alpha'})],
\\
T_{\alpha\beta\alpha'\beta'}&=&\sum_{\lambda_1\lambda_2\lambda_3\lambda_4}
\langle\lambda_1\lambda_2|v|\lambda_3\lambda_4\rangle
[\delta_{\alpha\lambda_1}
C_{\lambda_3\lambda_4\beta\alpha'\lambda_2\beta'}
+\delta_{\beta\lambda_2}C_{\lambda_4\lambda_3\alpha\beta'\lambda_1\alpha'}
\nonumber \\
&-&\delta_{\alpha'\lambda_3}C_{\alpha\lambda_4\beta\lambda_1\lambda_2\beta'}
-\delta_{\beta'\lambda_4}C_{\beta\lambda_3\alpha\lambda_2\lambda_1\alpha'}],
\label{T}
\\
I(\alpha\beta\gamma\alpha'\beta'\gamma')
&=&\sum_{\lambda_1\lambda_2\lambda_3}\{
\langle\alpha\lambda_3|v|\lambda_1\lambda_2\rangle
(n_{\gamma\lambda_3}C_{\lambda_1\lambda_2\beta\alpha'\beta'\gamma'}
-n_{\beta\lambda_3}C_{\lambda_1\lambda_2\gamma\alpha'\beta'\gamma'}
+C_{\lambda_1\lambda_2\alpha'\beta'}C_{\beta\gamma\gamma'\lambda_3}
\nonumber \\
&-&C_{\lambda_1\lambda_2\alpha'\gamma'}C_{\beta\gamma\beta'\lambda_3}
+C_{\lambda_1\lambda_2\beta'\gamma'}C_{\beta\gamma\alpha'\lambda_3})
\nonumber \\
&+&\langle\alpha\lambda_3|v|\lambda_1\lambda_2\rangle_A
[n_{\lambda_1\alpha'}C_{\lambda_2\beta\gamma\beta'\gamma'\lambda_3}
-n_{\lambda_1\beta'}C_{\lambda_2\beta\gamma\alpha'\gamma'\lambda_3}
+n_{\lambda_1\gamma'}C_{\lambda_2\beta\gamma\alpha'\beta'\lambda_3}
\nonumber \\
&+&C_{\lambda_2\beta\gamma'\lambda_3}C_{\lambda_1\gamma\alpha'\beta'}
+C_{\lambda_2\gamma\alpha'\lambda_3}C_{\lambda_1\beta\gamma'\beta'}
+C_{\lambda_2\gamma\beta'\lambda_3}C_{\lambda_1\beta\alpha'\gamma'}
-C_{\lambda_2\gamma\gamma'\lambda_3}C_{\lambda_1\beta\alpha'\beta'}
\nonumber \\
&-&C_{\lambda_2\beta\alpha'\lambda_3}C_{\lambda_1\gamma\gamma'\beta'}
-C_{\lambda_2\beta\beta'\lambda_3}C_{\lambda_1\gamma\alpha'\gamma'}
\nonumber \\
&+&n_{\gamma\lambda_3}(n_{\lambda_1\alpha'}C_{\lambda_2\beta\beta'\gamma'}
-n_{\lambda_1\beta'}C_{\lambda_2\beta\alpha'\gamma'}
+n_{\lambda_1\gamma'}C_{\lambda_2\beta\alpha'\beta'})
\nonumber \\
&-&n_{\beta\lambda_3}(n_{\lambda_1\alpha'}C_{\lambda_2\gamma\beta'\gamma'}
-n_{\lambda_1\beta'}C_{\lambda_2\gamma\alpha'\gamma'}
+n_{\lambda_1\gamma'}C_{\lambda_2\gamma\alpha'\beta'})
\nonumber \\
&+&n_{\lambda_1\alpha'}(n_{\lambda_2\beta'}C_{\beta\gamma\gamma'\lambda_3}
-n_{\lambda_2\gamma'}C_{\beta\gamma\beta'\lambda_3})
+n_{\lambda_1\beta'}n_{\lambda_2\gamma'}C_{\beta\gamma\alpha'\lambda_3}]\},
\\
J(\alpha\beta\gamma\alpha'\beta'\gamma')
&=&
\sum_{\lambda_1\lambda_2}[\langle\alpha\beta|v|\lambda_1\lambda_2\rangle_A
(n_{\lambda_1\alpha'}C_{\lambda_2\gamma\beta'\gamma'}
-n_{\lambda_1\beta'}C_{\lambda_2\gamma\alpha'\gamma'}
+n_{\lambda_1\gamma'}C_{\lambda_2\gamma\alpha'\beta'})
\nonumber \\
&+&\langle\alpha\beta|v|\lambda_1\lambda_2\rangle
C_{\lambda_1\lambda_2\gamma\alpha'\beta'\gamma'}].
\end{eqnarray}
The matrix
$B_{\alpha\beta\alpha'\beta'}$ does not contain $C_{\alpha\beta\alpha'\beta'}$ and may be called 
the Born term, whereas $P_{\alpha\beta\alpha'\beta'}$ and $H_{\alpha\beta\alpha'\beta'}$
contain $C_{\alpha\beta\alpha'\beta'}$ and describe
particle-particle (and hole-hole) and particle-hole 
correlations to infinite order, respectively. The last term on the right-hand side of eq. (\ref{F2}) 
expresses the contribution of the three-body correlation matrix. The matrix
$J(\alpha\beta\gamma\alpha'\beta'\gamma')$ includes particle-particle (and hole-hole) correlations,
while $I(\alpha\beta\gamma\alpha'\beta'\gamma')$ contains both particle-particle and 
particle-hole correlations. 

\end{document}